# Mechanical Properties of Gradient Nano-Gyroid Cellular Structures: A Molecular Dynamics Study

Rui Dai [1]; Dawei Li [2*]; Yunlong Tang [3*]

1. Department of Mechanical Engineering, School of Engineering for Matter, Transport and Energy, Arizona State University, Tempe, AZ 85281, United States

2. School of Mechanical Engineering, Nanjing University of Science and Technology, Nanjing, Jiangsu 210094, China

3. Department of Mechanical & Aerospace Engineering, Monash University, Clayton Victoria 3800, Australia

Email:   davidlee@nuaa.edu.cn ; yunlong.tang1@monash.edu

*. Corresponding author

**Graphical abstract**

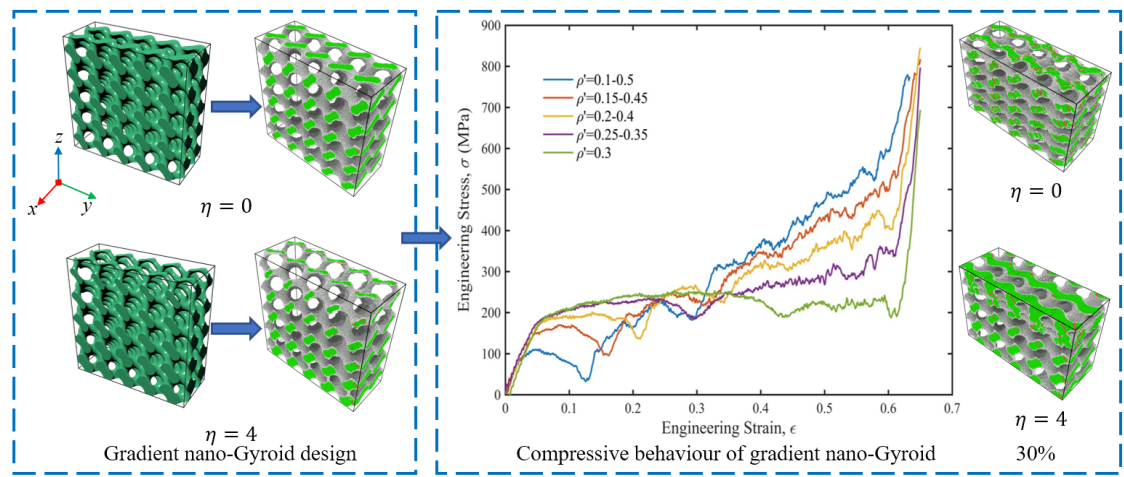

**Abstract:**

Advanced manufacturing (AM) technologies, such as nanoscale additive manufacturing process, enable the fabrication of nanoscale architected materials which has received great attention due to their prominent properties. However, few studies delve into the functional gradient cellular architecture on nanoscale. This work studied the gradient nano-Gyroid architected material made of copper (Cu) by molecular dynamic (MD) simulations. The result reveals that, unlike homogeneous architecture, gradient Gyroid not only shows novel layer-by-layer deformation behaviour, but also processes significantly better energy absorption ability. Moreover, this deformation behaviour and energy absorption are predictable and designable, which demonstrate its highly programmable potential.

## 1. Introduction

Nanoscale architected materials are utilized as the building blocks for macro matters, which have demonstrated their high potential applications[1]. Most of these materials are the inspiration from nature and reflects nature's vast majority of functional structures. Natural material architectures commonly process porous and hierarchical structures and are evolved to have compelling multifunctional properties[2]. Among them, the single Gyroid minimum geometry, a porous, interconnected with the cubic symmetry material is one intriguing representative geometry and a member of triply periodic minimal surface (TPMS) family.[3] One example of natural Gyroid structure can be found on the wing of butterfly, as shown in Figure 1(a), which is fascinating subject to the increasing interests in the structural color, photonics and optics[1, 4-7]. As an architecture porous structure and being simultaneous lightweight and strong, it can potentially be made as a thin film or a battery electrode applied in electronic sectors[8] as well as achieving high impact energy absorption employed in industrial applications, such as automobile and aerospace sectors[9] as well as personnel protection[10] or package protection.

To further understand the geometry and biomimetic from nature, several different types of Gyroid based architecture cellular materials are fabricated by additive manufacturing (AM) technologies [11, 12]. For example, Diab et al.[11] fabricated polymeric Gyroid cellular and found that such structure had comparable mechanical properties to other TMPS structures. With the advances of AM techniques, examples like high-resolution AM and laser lithography techniques, it is feasible to fabricate 3D Gyroid cellular with feature resolution on the scale of few to few hundreds nanometers. For example, Gan et al.[4] demonstrated that using optical two-beam lithography technique can replicate Gyroid photonic nanostructures in the *butterfly Callophrys Rubi* up to several hundred nanometers and potentially aid in the investigation of biomimetic structures.

The rapid developments in fabricating nanolattices have been continuously extending the material space and redefining the mechanical limit[2]. Given that the size of the material structure decreases below micro-scale, many architectural cellular materials which made of single-crystalline metal exhibit the size effect. It can tremendously alter the mechanical and electrical properties compared to those cellular materials fabricated on the macroscale. Thus, the study of nanosized architectural cellular materials is essential. With the size effect, nanoscale cellular materials exhibit unique and unparalleled mechanical properties, known as "the smaller the stronger" e.g. high deformability[2, 13], the superior recoverability with brittle material[14] and ultrahigh effective strength[13]. Moreover, numerous studies have demonstrated the exceptional mechanical properties of nanoscale Gyroid (nano-Gyroid) and Gyroid-like structures. For example, Park et al.[15] shows the nano-Gyroid yields an unusually high ratio of shear modulus to young's modulus, which is twice larger than conventional bulk materials. Besides, he found the young's ($E$) and shear ($\mu$) modulus of nanoscale TPMS structures change with the relative density and size at different rates for different morphologies[16]. Qin et al.[17] investigated the mechanics of fundamental Gyroid graphene structure and reveals that although 3D graphene has exceptional mechanical strength and stiffness, its mechanical properties decrease with density much faster than those of polymer foams. What's more, various experiments and simulations show nanoscale Gyroid-like TPMS structures display a relatively uniform stress distribution under compression leading to a stable collapse mechanism since they process no discontinuities and therefore avoid stress concentration during deformation[10, 11, 18, 19]. Despite those studies mentioned above, to the author's best knowledge, nevertheless, there is limited research concerned with the mechanical properties of nanoscale functionally graded cellular structures, and their impact energy properties is yet to be studied.

Single crystal Copper (Cu) lattice composed of nanotwinned structural elements have significant improved mechanical strength compared to the equivalent bulk materials. However, it's still unknown how the size effect, which becomes critical under 100nm scale, impacts the mechanical properties of nanolattices. Single crystal copper (Cu) lattice composed of nanotwinned structural elements have significantly improved mechanical strength compared to the equivalent bulk material[22]. To provide a reliable understanding of the mechanical properties of nanoscale functionally graded Gyroid structures, we perform MD simulations on a group of nanoscale functionally graded Gyroid structures made of Cu. Interestingly, we find, unlike the homogeneous Gyroid structure, the energy absorption ability linearly dependents on the gradient level. However, noteworthy, when the gradient distinction is large enough, surface effects may become dominant resulting the nonlinear effect. Moreover, the layer by layer collapse behavior reflecting a highly programmable potential, e.g. stiffness and strength. The result of this work will be instrumental in understanding the mechanical response of gradient Gyroid cellular on nanoscale and developing new class of stable nanoscale metamaterials.

## 2. Methods

### 2.1 Uniform and Gradient Gyroid Generation

Gyroid is one of the most widely studied types of triply periodic minimal surfaces (TPMS), which was discovered in the 1970s by NASA scientist Alan Schoen and designed for lightweight high-strength new materials. Additionally, the Gyroid structure is stably present in nature on nanoscale, such as the microstructure of butterfly wings (as shown in Figure 1a). Gyroid surface can use a triply periodic function to express its spatial form (in Figure 1b):

$$F_G(x,y,z) = \sin\left(2\pi\frac{x}{L}\right) \times \cos\left(2\pi\frac{y}{L}\right) + \sin\left(2\pi\frac{y}{L}\right) \times \cos\left(2\pi\frac{z}{L}\right) + \sin\left(2\pi\frac{z}{L}\right) \times \cos\left(2\pi\frac{x}{L}\right) - t$$
(1)

where $L$ is the cubic unit cell edge length, and the level parameter $t$ is a variable and determines the volume fractions in the regions separated by the surface. Then, when the surface is closed, the Gyroid structure can be established (in Figure 1c). Different volume fractions of Gyroid structure can be obtained by varying the parameter $t$, and its relationship to relative density $\rho^*$ is shown in Figure 1d. As shown in Figures 1e and 1f, the uniform and gradient Gyroid structure, respectively, have a size parameter of 10×43×43nm and average density 0.3. And the maximum density minus the minimum density divided by 0.1 is defined as the structure gradient $\eta$. For example, the Nano-Gyroid structure gradient of uniform density is 0 and density range 0.1 ~ 0.5 is 4, respectively.

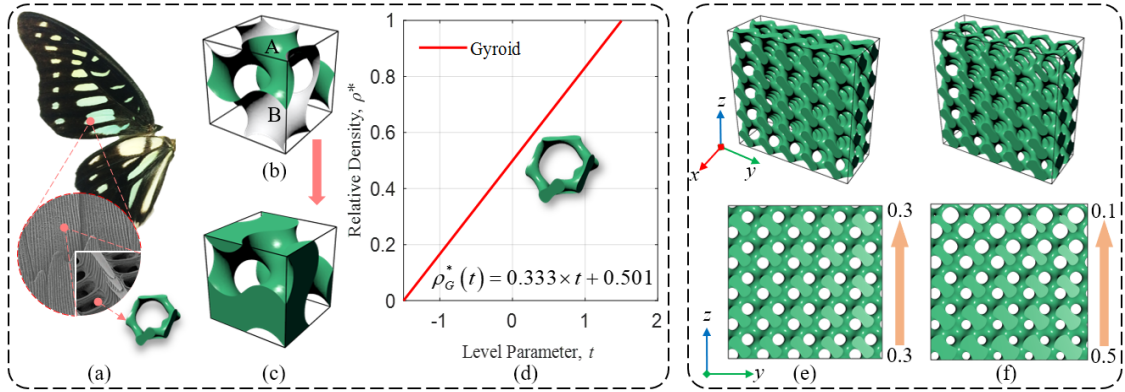

**Figure 1.** Nano-scale Gyroid structure modeling and compression deformation process: (a) microstructure of butterfly wings; Gyroid surface (b) close to be solid Gyroid structure (c); (d) is the relationship between level parameter $t$ and relative density; (e) and (f) are the nano-Gyroid structures with gradients 0 and 4, respectively.

**2.2 Molecular Dynamics (MD) Simulation for Nano-Gyroid Structure**

MD simulations are performed in this work using LAMMPS package[20] with embedded-atom (EAM) potential parametrized by Mishin et al.[21] to study the gradient Gyroid cellular structure and investigate their elastic response. The parameterization of the adopted EAM potential is obtained using the ab-initio electronic calculations. Moreover, the potential utilized is proved to be capable of accurately capturing the essential deformation behavior after analyzing the stress-strain relation, atomic relaxation and $\gamma$ surface for {111} shear. It makes the EAM potential feasible to study large atomic systems compared with ab-initio electronic calculations [22].

To elucidate the mechanical properties of the nano-Gyroid structure, all the molecular dynamic simulations are performed with structure gradient ranging from 0 to 4. To construct the nanolattice, FCC bulk copper (edge length around 40nm) with over 2 million atoms is generated as single crystal oriented all along the <100> direction. Subsequently, extra atoms are removed according to Equation (1) resulting the uniform Gyroid and gradient nano-Gyroid cellular structures containing around 840,000 atoms and a relative density 0.1 to 0.5, respectively. Periodic boundary conditions (PBC) are applied along all three directions of the nanolattice cell. Provided that lattice deformation behaviors along *z*-direction are to be considered, only two units are created along *x*-direction. Initially, we performed the molecular dynamics simulations within the Isothermal-Isobaric (NPT) ensemble for 800ps with a time step of 1 fs. Such treatment is supposed to relax the system while letting the structure reaches its equilibrium state. During the equilibrium process, the total temperature was kept at 300 K and the pressure along all three axial directions were kept at zero. After that, the equilibrated structure was compressed along *z*-direction with a compression strain up to 65% of the original length. The deformation controlled uniaxial compression processes a strain rate of $10^9 \, s^{-1}$ and the pressure component perpendicular to the compression direction was controlled maintaining the uniaxial compression condition. The method to calculate the atom stress is described below. The atomic stress tensor $S_{ij}^\alpha$ at atom $\alpha$ is calculated by the following equation, where $i$ and j act as *x*, *y* or *z* indicating 6 different components of the stress tensor.

$$S_{ij}^\alpha = \frac{1}{2} m^\alpha v_i^\alpha v_j^\alpha + \sum_{\beta=1}^{n} r_{\alpha\beta}^j f_{\alpha\beta}^i$$

$m^\alpha$ and $v^\alpha$ are the mass and velocity of atom $\alpha$; $r_{\alpha\beta}$ and $f_{\alpha\beta}$ are the distance and force between atoms $\alpha$ and $\beta$. Once the stress tensor is obtained, the equivalent stress $\sigma$ of nano-Gyroid structure is calculated based on von Mises stress described as

$$\sigma = \sqrt{\frac{1}{2}\left[\left(\sigma_{xx} - \sigma_{yy}\right)^2 + \left(\sigma_{yy} - \sigma_{zz}\right)^2 + (\sigma_{zz} - \sigma_{xx})^2 + 6\left(\sigma_{xy}^2 + \sigma_{yz}^2 + \sigma_{zx}^2\right)\right]}$$

Here, $\sigma_{ij} = \frac{1}{V}\sum_{\gamma=1}^{n} S_{ij}^\gamma$, where $V$ is the volume of the system.

3. Results and Discussion

3.1 Mechanical Properties of Homogeneous and Gradient Gyroid

We performed MD simulations of compression of the nano-Gyroids on different relative densities up to the strain of 0.65, as shown in Fig. 2. At around 60% deformation, the Gyroid nanolattices begin to fail and lose their loading capacity. The compression stress-strain curve of nano-Gyroid structures with gradient densities is shown in Fig. 2 (a). In Fig 2. (a) depending on different relative density level the compression response varies showing a distinct mechanical behavior with macroscale bulk Cu lattice structures. To be specific, initially an approximate linear behavior is observed with compression strain ranging from 2.0% to 4.3%, followed by the distinct plastic yielding then another linear region and plastic yielding stage. This alternative behavior is never observed in pure bulk crystals. After each yielding, we observed a plateau in the flow stress followed by failure via layer densification. However, for the uniform Gyroid model ($\eta = 0$), there is no obvious plateau stress region. Fig. 2 (b) presents schematic diagrams of the von Mises stress distribution along z-direction for $\eta = 0$ and $\eta = 4$ model respectively. The uniform Gyroid fails all over the structure and almost every layer of structure is simultaneously compressed and deformed throughout the process before being densified, while the graded one fails only layer by layer providing a continuous decay in mechanical stiffness, which indicates the potential of the gradient nano-Gyroid structure performing as a programmable architecture to process larger deformation to be applied in broad areas, e.g. energy absorption, damage tolerance and recoverability.

Another thing we find is the young's modulus, for traditional structure, the effective young's modulus is defined with the gradient of the first 2% to 4% deformation region. However, unlike standard uniform structure, for the gradient architecture, there is no linear deformation part no matter how small the initial region is, which originates from the inhomogeneous nature. The inset image in Fig. 2 (a) is a representative stress-strain response for gradient 2 model. Herein the young's modulus is derived from the largest gradient within the starting 2% to 4% deformation region. Among the five models, $\eta = 2$ model (relative density 0.15-0.45) processes the strongest stiffness among others, which cannot be explained with Gibson-Ashby cellular material theory[23] and need further study.

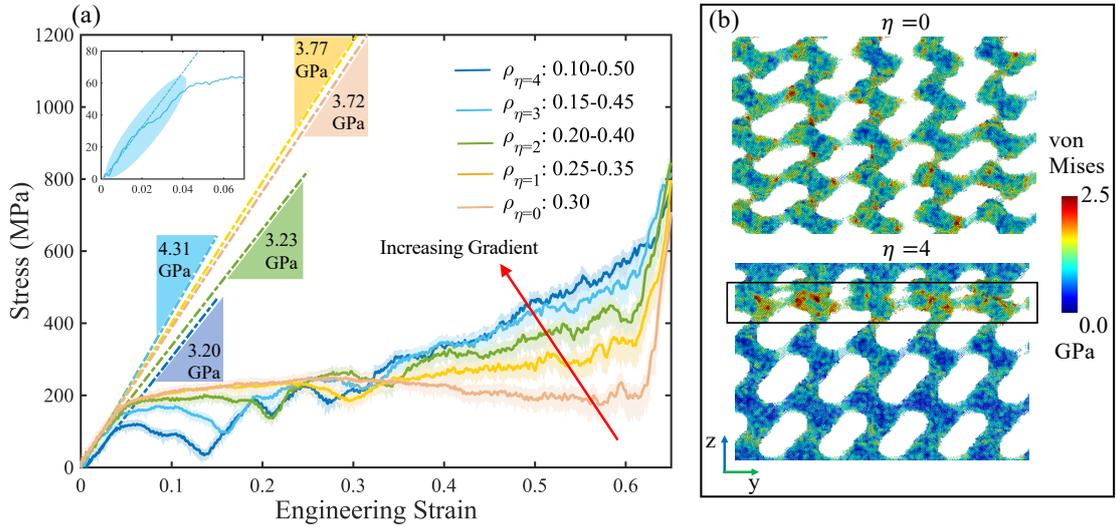

**Figure 2.** (a) Compressive stress-strain curves of <100> Gyroid lattice structures with different gradients. (b) Schematic diagram of the atomic von Mises stress distribution for $\eta = 0$ and $\eta = 4$, respectively. The red areas indicate the yielding regions. A black rectangular is marker of the collapse region on top layers for $\eta = 4$ model.

To connect the mechanical response with atomic-scale deformation, we provide the compression deformation of $\eta = 0$ and $\eta = 4$ structures at several levels of overall stain, from 0.02 to 0.3 shown in Fig.3. The uniform and the gradient nano-Gyroid cellular structures exhibit completely different deformation response. For the uniform Gyroid model, the compression process is divided into two stages (I) elastic stage (II) plastic stage and (III) densification.

Unlike the uniform structure, gradient Gyroid deformation can be divided into five stages containing: (I) (II) (III) elastic stage and plastic stage, (IV) densification. The densification in larger graded structure leads to a rapid increase in stress above the yield stress, which occurs due to the continuous densification within the unit cell as compared to uniform structure. Noteworthy, for the $\eta = 4$ model, there is buckling near the 0.1 density region which is not observed in the uniform structure due to the small aspect radio to be approximate as a thin truss.

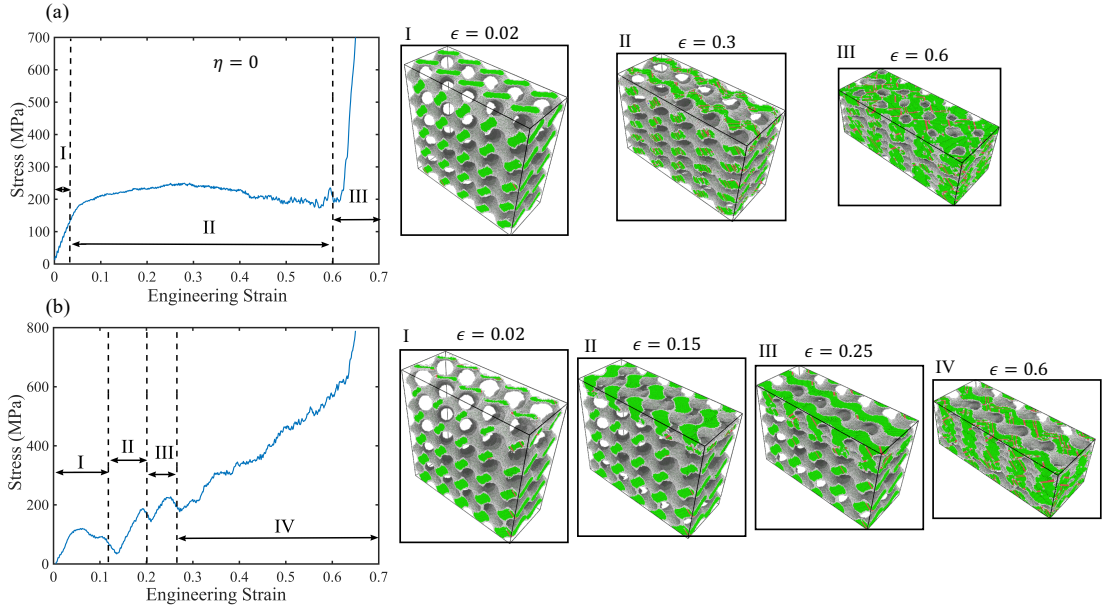

**Figure 3.** Engineering stress-strain curve of Gradient 0 and Gradient 4 model with <100> nanolattice (a) and (b).

### 3.2 Energy Absorption of Nanoscale of Uniform and Gradient Gyroid

From the stress-strain curve of Figure 2a, it can be seen that the nano-Gyroid with a gradient from 0 to 4 all reached a dense state. Moreover, the energy absorption per unit volume with a densification strain $\varepsilon_d$ can be calculated according to formula (2).

$$W_v = \int_0^{\varepsilon_d} \sigma(\varepsilon) d\varepsilon \tag{2}$$

where $W_v$ is the cumulative energy absorption per unit volume for nano-Gyroid structures, $\varepsilon$ is the strain, and $\sigma(\varepsilon)$ is the stress related to $\varepsilon$ during the compression simulations. From Figure 4 (a), it can be seen that the smaller the structural gradient is, the longer the linear region of the curve is.

With the increase of the structural gradient, the more nonlinear the curvature of the relationship between energy absorption and strain is. This is because the structure with gradient 0 has a longer plastic strain region. Additionally, as the structural gradient increases, its dense cumulative absorption capacity also gradually increases. The results show that the larger the gradient of nano-Gyroid structure, the higher the energy absorption capacity. Therefore, the gradient structure can provide greater strain protection before densification. Moreover, the numerical value shows that the cumulative energy absorption value of the nano-Gyroid structure with a gradient of 4 is about 37.4% higher than that of a uniform structure with a gradient of 0 (Table. 1). Furthermore, from Figure 4 (b), it can be seen that the stress of the structure with small gradient increases rapidly when the energy absorption increases to a certain value. For structures with larger gradients, the stress gradually increases with the energy absorption value, which is more conducive to buffer energy absorption.

Besides, as shown in Figure. 5, the relationship between the gradient of the nano-Gyroid structure and the cumulative energy absorption value is shown. It can be seen from the fitted result that, there is an approximately linear relationship between the gradient change and the cumulative energy absorption. Therefore, a linear fitting equation can be established to predict the cumulative energy absorption corresponding to different gradients.

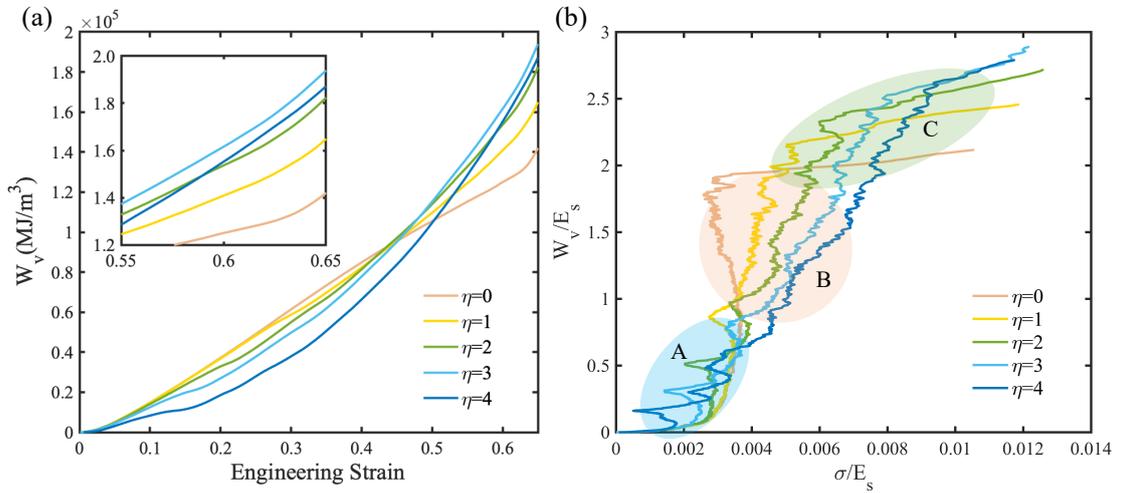

**Figure 4.** (a) Cumulative energy absorption per unit volume versus strain curves and (b) normalized energy absorption of nano-Gyroid structures with gradients 0 and 4, respectively.

**Table 1.** Densification strains $\varepsilon_d$ and energy absorbed per unit volume at densification $W_v$ for the gradient nano-Gyroid structures.

| Type | $\eta = 0$ | $\eta = 1$ | $\eta = 2$ | $\eta = 3$ | $\eta = 4$ |
|---|---|---|---|---|---|
| Strain $\varepsilon_d$ | 0.615 | 0.608 | 0.606 | 0.615 | 0.634 |

| Energy absorbed $W_v$ ($MJ/m^3$) | 127936.2 | 143601.8 | 156081.1 | 169519.6 | 175807.1 |

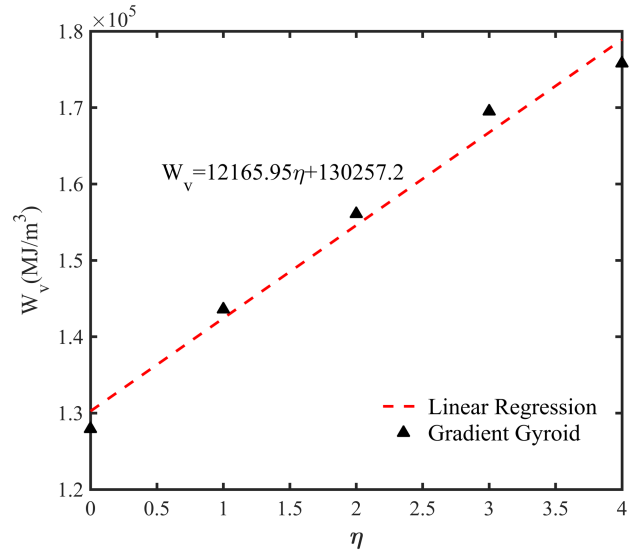

**Figure 5.** Predicting the relationship between gradient and energy absorption.

## 4. Conclusion

In summary, nanoscale Gyroid with different gradients were designed and their mechanical properties were characterized by MD simulation. Compared with the overall compressive deformation of a homogeneous nano-Gyroid structure, the gradient structure is compressed hierarchically and destroyed layer by layer. Furthermore, the results show that as the gradient increases, the overall energy absorption value of the structure increases. Therefore, the relationship between the gradient level and energy absorption is established to predict energy absorption.

**Declaration of Competing Interest**

The authors declare that they have no known competing financial interests or personal relationships that could have appeared to influence the work reported in this paper.

**Acknowledgment**

This work is supported by Financial Support from the Program of China Scholarships Council under Award Number 201706830039.